\documentclass[preprint,aps,showpacs]{revtex4}
\usepackage{graphicx}
\usepackage{amsmath}
\usepackage{amsfonts}
\usepackage{epsfig}
\begin{document}

\title{
Nuclear halo and the coherent nuclear interaction}
\author{{\bf Renzo Alzetta}, {\bf Giuseppe Liberti\footnote{Corresponding author.
E-mail address: liberti@fis.unical.it}}, {\bf Rosa L. Zaffino}}
\affiliation {Dipartimento di Fisica, Universit\`a della Calabria\\
INFN, Gruppo Collegato di Cosenza, I-87036 Rende (CS), Italy}
\date{\today}
\begin{abstract}
The unusual structure of $^{11}Li$, the first halo nucleus found,
is analyzed by the Preparata model of nuclear structure. By
applying Coherent Nucleus Theory, we obtain an interaction
potential for the halo-neutrons that rightly reproduces the
fundamental state of the system.
\end{abstract}
\bigskip
\maketitle

\section{Introduction}
The understanding of halo nuclei \cite{Tani} is one of the issues
of nuclear research still awaiting a satisfactory explanation. The
nucleus ${}^{11} Li$ is the first observed and the most
interesting case of a two neutrons halo (${}^{9} Li + n + n$) and
different kinds of experiment \cite{boh} have been performed to
investigate its structure. The generally accepted picture of a
halo nucleus describes it by a core nucleus surrounded by loosely
bound valence neutron which tunnel with significant probability
into a region outside the core potential. Many theoretical efforts
have been carried out to describe this Borromean system where the
three-body system is bound and its all binary subsystem are
unbound, but these attempts with the nature of the ${}^{9} Li+n$
interaction that is not exactly known.

In this paper we wish to show that the unusual properties of halo
nuclei can be explained in the framework of the Coherent Nucleus
Theory proposed more the ten years ago by Giuliano Preparata
\cite{Preparata1,Preparata2}. This theory which lays at the
foundation of the nuclear Shell Model has been applied to several
problems of nuclear physics, namely: EMC nuclear effect
\cite{Preparata3}, Coulomb sum rule in quasi-elastic
electron-nucleus scattering \cite{Ractliffe}, deep-inelastic
scattering at low $x$ \cite{Preparata4}, hypernuclear interactions
\cite{Preparata5} and decays \cite{ipdeca}, low-energy
photoabsorption in nuclei \cite{foto}, neutron stars
\cite{neustar}. According to this approach, inside a nucleus the
nucleons are involved in a laser-like process whose two levels,
strongly coupled to the pion field, are the $N(940)$ and the
$\Delta(1232)$. The solutions of the coherence equations for this
$N \Delta \pi$ coupled system, at the resonant $\pi$-mode for
which $\omega_{\vec q}\equiv \omega_{o}= m_{\Delta}-m_{N}=292$ MeV
and $q \equiv |\vec q\,|=256$ MeV, are characterized by
time-independent amplitudes and phases that vary linearly with
time. In this way, the laser process produces a coherent
$\pi$-condensate, characterized by its well-defined phase relation
with the N-$\Delta$ system, which is spread out throughout the
spatial region where the collective "N-$\Delta$ current" is
localized, i.e. within the nucleus.

The most stable configuration is reached in a spatial region of
radius $R_{CD}\simeq 4.2$ fm called Coherence Domain (CD) and
within a single CD the resulting $p$-wave $\pi$-field is given by
($i=1,2,3$ is the isospin index)
\begin{equation}
    \phi_{i}(\vec x ,t)= 8\pi \sqrt{\frac{\rho}{2\omega_{o}}}(\hat x
    \cdot \vec{\alpha}_{i}) j_{1}(q r) \sin (\omega_{r}t)  \quad (r<R_{CD})
    \label{pionfield}
\end{equation}
where $\rho$ is the nuclear density, $\vec{\alpha}_{i}$ is the
$\pi$-amplitude
\begin{equation}
 \sum_{i} |\vec{\alpha}_{i}|^{2} \simeq
(0.3)^{2}
\label{alfa}
\end{equation}
and
\begin{equation}
\omega_{r}= \omega_{o}(1-\dot{\phi})\simeq 100\,{\mathrm{MeV}}
\label{omega}
\end{equation}
is the "renormalized frequency" ($\dot{\phi}$ is the pion phase
velocity) inside the nuclear medium. We should emphasize that this
pion condensation, that is generated via long-range coherent
hadronic forces, is totally unrelated to the static incoherent
pion condensate proposed by Migdal \cite{Migdal} and predicted at
densities far from normal nuclear density \cite{Weise}. As a
fundamental result, the total energy of the coherent state is
lowered and the average energy gain per particle, at nuclear
matter density $\rho_{o}\simeq 0.166$ fm$^{-3}$, is about $60$
MeV, which well represents the depth of the self-consistent
nuclear potential.

A coherent evolution of the nuclear dynamics is done also for
light nuclei \cite{foto} if it is possible to match
Eq.(\ref{pionfield}) with the solution of the free-field equation
($\hbar=c=1$)
\begin{equation}
    (\Box +m_{\pi}^{2})\phi_{i}(\vec x ,t)=0
    \label{free-pi}
    \end{equation}
valid outside the nucleus $(r>R_{A}\simeq r_{o}A^{\frac{1}{3}})$,
whose $p$-wave solution is given by
\begin{equation}
\phi_{i}(\vec x , t)=(\hat x \cdot \vec{A}_{i})k_{1}(\lambda
r)\sin(\omega_{r}t) \quad (r>R_{A}) \label{out-pion}
\end{equation}
where
\begin{equation}
    \vec{A}_{i}=8\pi \sqrt{\frac{\rho}{2\omega_{o}}}
    \frac{j_{1}(qR_{A})}{k_{1}(\lambda R_{A})} \vec{\alpha}_{i}
    \label{Acappa}
    \end{equation}
and $\lambda$ determined by joining Eq.(\ref{out-pion}) together
with its first radial derivative to the inner $\pi$-field of
Eq.(\ref{pionfield}). A simple calculation provides for a critical
radius $R_c\simeq 2.42$ fm below which there can be no
exponentially decaying solution of (\ref{free-pi}). The radius of
${}^9 Li$ is approximately $R_A\simeq 2.5$ fm and we obtain
$\lambda \simeq 62$ MeV.

We posses now all the ingredients needed to analyze a possible
mechanism for explaining the basic properties of the halo nuclei
in terms of interaction between the extra neutrons and the
evanescent tail of the pion condensate (\ref{out-pion}).

\section{Model and method of calculation}
In our approach to determine the ground state of the halo nucleus
we assume that the extra neutrons interact with the core-nucleus
through their coupling to the {\it evanescent} tail of its
coherent pion field $\pi_{c}$. The virtual dispersive interaction
potential for the basic process $\pi_{c}+n\rightarrow \pi_{c}+n $
is calculated by applying second order perturbation theory:
\begin{equation}
    V_{n}(\vec{x})=-\frac{i}{4 m_{n}}\int_{-\infty}^{+\infty}dt
    \int d^{3}{\xi} \langle n|T\left[ H_{I}\left(\vec{x}+\frac{\vec{\xi}}{
    2},\frac{t}
     {2}\right)H_{I}\left(\vec{x}-\frac{\vec{\xi}}{2},-\frac{t}
     {2}\right)\right]|n\rangle
    \label{potint1}
    \end{equation}
where $|n\rangle$ is the ground state we search for and the
interaction hamiltonian $H_{I}$ will be given explicitly later.

Inserting into Eq.(\ref{potint1}) a complete sum over intermediate
state and re-arranging the time-ordered product we obtain:
\begin{equation}
    V_{n}(\vec{x})=-\frac{i}{2 m_{n}}\int_{0}^{+\infty}dt
    \int d^{3}{\xi} \sum_{N}\langle n| H_{I}\left(\vec{x}+\frac{\vec{\xi}}{
    2},\frac{t}{
     2}\right)|N\rangle\langle N|H_{I}\left(\vec{x}-\frac{\vec{\xi}}{
    2},-\frac{t}
    {2}\right)|n\rangle
    \label{potint2}
    \end{equation}
For our low-energy calculation the intermediate state are the
nucleon themselves and the interaction Hamiltonian that we assume
is the usual non relativistic reduction of the pseudo-scalar $\pi
NN$ coupling:
\begin{equation}
    H_{\pi NN}= ig \bar{N}\gamma_{5}\vec{\tau}\cdot \vec{\pi} N
    \label{hamint}
    \end{equation}
with ${g^{2}/{4\pi}}\simeq 14.3$.\\
In non-relativistic limit the matrix element is:
\begin{equation}
   \langle{n}|H_{I}(\vec{x},t)|{N}\rangle=g(\vec{\sigma}\cdot\vec{k})[\vec{\tau}\cdot\vec{\phi}(\vec{x})]e^{i\vec{k}\cdot\vec{x}}
   e^{-i(E_N-m_{n}t)}\sin{(\omega_{r}t)}
  \label{matrelmt}
  \end{equation}
  where $\vec{\phi}(\vec{x})$ is the $\pi$-field of
Eq.(\ref{out-pion}), $\vec{k}$ is the momentum of the intermediate
nucleon, $\vec{\sigma}$ and $\vec{\tau}$ are the spin and the
isospin operators respectively. Performing the necessary algebra
and integrating over the time, we have:
\begin{eqnarray}
    V_{n}(\vec{x})&=&-\frac{g^{2}} {4m_{n}} \int \frac{d^{3}k} {(2\pi)^{3}}
    k^{2}\left[ \frac{1}  {k^{2}}-\frac{1} {2}\left(
    \frac{1}{k^{2}+2m_{N}\omega_{r}}\right)+\left(
    \frac{1}{k^{2}-2m_{N}\omega_{r}}\right)\right]\nonumber\\
    &\times&\sum_i\int d^{3}{\vec{\xi}} \phi_{i}\left(\vec{x}+\vec{\xi}/2\right)
    \phi_{i}\left(\vec{x}-\vec{\xi}/2\right) e^{i\vec{k}\cdot
    \vec{\xi}}
    \label{potint3}
    \end{eqnarray}
Inserting the explicit expression for the pion field and
performing the integration over the variable $\vec{\xi}$ we
obtain, taking the limit $|\vec{\xi}| \ll |\vec{x}|$, the
following result
\begin{eqnarray}
  V_{n}(\vec{x})&\simeq&-\frac{g^2}{4m_{n}}\left(\frac{4\pi
 r}{\lambda}\right)^{\frac{3}{2}}\frac{e^{-2\lambda r}}{\lambda^{2}r^{2}}
 \left(1+\frac{1}{\lambda r}\right)^{2}\sum_i(\hat{x}\cdot\vec{A_{i}})^2\nonumber\\
 &\times&\int\frac{d^{3}{k}}{({2\pi})^3}e^{-\frac{r}{\lambda}\vec{k}^2}
  \left[1-\frac{\vec{k}^2}{2}\left(\frac{1}{\vec{k}^2+2m_N\omega_{r}}
+\frac{1}{\vec{k}^2-2m_N\omega_{r}}\right)\right]
 \label{eq:potint5}
 \end{eqnarray}
where $r=|\vec{x}|$. The integration over $|\vec{k}|$ is extended
to all those values for which the exponential in the above
expression doesn't make the integrand function vanishing,
therefore the significant region for the integration is such that
$R_{halo}k^{2}/\lambda < 1$. For this set of values
  we have $k\ll\sqrt{2m_N\omega_{r}}$ and
\begin{equation}
    \int{d^{3}{k}}e^{-\frac{r}{\lambda}\vec{k}^2}
\vec{k}^2\left(\frac{1}{\vec{k}^2+2m_N\omega_{r}}+\frac{1}{\vec{k}^2-2m_N\omega_{r}}\right)\simeq
0 \label{intk}
\end{equation}
that allows us to rewrite the potential as:
\begin{equation}
V_{n}(\vec{x})\simeq-\frac{g^2}{4m_{n}} \frac{e^{-2\lambda
r}}{\lambda^2 r^2}\left(1+\frac{1}{\lambda
r}\right)^2\sum_i(\hat{x}\cdot\vec{A_{i}})^2, \quad r> R_A
 \label{eq:potint7}
\end{equation}

The process $\pi_{c}+[nn]\rightarrow \pi_{c}+ [nn]$, actually the
one involving the neutron pair in $^{11}Li$, is a few-body problem
whose solution requires some approximations. The interaction
Hamiltonian is
\begin{equation}
    H_{I}=i\frac{g}
    {2m_{n}}\sum_{\lambda=1}^{2}\vec{\tau}_{\lambda}(\vec{\sigma_{\lambda}}\cdot
    \vec{k}_{\lambda})\vec{\phi}(\vec{x}_{\lambda})
    \label{2hamin}
    \end{equation}
where $\vec{x}_{\lambda}$ and $\vec{k}_{\lambda}$ stand for
position and momentum of the $\lambda^{th}$ neutron
$(\lambda=1,2)$ respectively.

A special case of interest that will be explored is that for which
the interaction Hamiltonian reduces to
\begin{equation}
     H_{I}=i\frac{g} {2m_{n}} [\tau_{1}^{k}(\vec{\sigma}_{1}\cdot
     \vec{k})+\tau_{2}^{k}(\vec{\sigma}_{2}\cdot \vec{k})]\phi_{k}(\vec{x})
     \label{2hamin2}
     \end{equation}
where
\begin{equation}
\vec{k}=\vec{k}_1+\vec{k}_2,\quad
\vec{x}=\frac{\vec{x}_{1}+\vec{x}_{2}}{2}
    \label{zeroapp}
    \end{equation}
At this point, with the adopted approximation, we can proceed as
the one extra neutron case. An intermediate result for the
potential is given by
\begin{eqnarray}
V_{2n}(\vec{x})&\simeq &-\frac{g^{2}}{2m_{n}}\left(\frac{4\pi r}
{\lambda}\right)^{\frac{3}{2}}\frac{e^{-2\lambda
r}}{\lambda^{2}r^{2}}
 \left(1+\frac{1}{\lambda r}\right)^{2}\nonumber\\
&\times &
\int\frac{d^{3}{k}}{(2\pi)^{3}}e^{-\frac{r}{\lambda}\vec{k}^{2}}
\left[\sum_i(\hat{x}\cdot\vec{A}_i)^{2}+(\vec{\sigma}_{1}\cdot\hat{k})(\vec{\sigma}_{2}\cdot\hat{k})
\tau_{1}^i
\tau_{2}^j(\hat{x}\cdot\vec{A}_i)(\hat{x}\cdot\vec{A}_j)\right]
\label{2eq:Vint8}
\end{eqnarray}
but, with respect to the case of one neutron, additional
considerations have to be done for the spin-isospin terms in the
above expression. When we consider the $n-n$ system in a singlet
spin state $S=0$ and in a triplet isotopic spin state $I=1$ we
have
\begin{equation}
\vec{\sigma}_{1} \cdot \vec{\sigma}_{2}=-3
\label{sigma}
\end{equation}
and
\begin{equation}
\chi^{\dagger}_{I=1}\tau^{i}_{1}\tau^{j}_{2}\chi_{I=1}=(\chi^{+}_{1})^{\dagger}
(\chi^{-}_{2})^{\dagger}
\tau^{i}_{1}\tau^{j}_{2}\chi^{+}_{1}\chi^{-}_{2}=\delta^{i3}\delta^{j3}
\label{tau}
\end{equation}
\begin{equation}
\tau_{1}^i
\tau_{2}^j(\hat{x}\cdot\vec{A}_i)(\hat{x}\cdot\vec{A}_j)=
(\hat{x}\cdot\vec{A}_3)^{2}=\frac{1}{3}\sum_i(\hat{x}\cdot\vec{A}_i)^{2}
\label{alfa}
\end{equation}
the last from symmetry conditions. The interaction potential is
finally given by
\begin{equation}
V_{2n}(\vec{x})\simeq -\frac{g^{2}}{3m_{n}}
 \left(\frac{R_{A}}{r}\right)^{4}
\left(\frac{1+\lambda r}{1+\lambda
R_{A}}\right)^{2}e^{-2\lambda(r-R_{A})}\sum_i(\hat{x}\cdot\vec{A_{i}})^2,
\quad r> R_A \label{eq:Vintfinalbis}
\end{equation}

\section{Results}

The effective Schr\"{o}dinger equation for ${}^9 Li+2n$ system is
given by
\begin{equation}
\left[-\frac{\nabla^{2}}{4
m_n}+V_{2n}(\vec{x})\right]\psi(\vec{x})=E_{2n}\psi(\vec{x}),\quad
r=|\vec{x}|>R_A \label{scroe}
\end{equation}
This equation can be solved for the state with $l=0$ observing
that
\begin{equation}
V_{2n}(\vec{x})_{l=0}\equiv
V_{2n}(r)=-V_0\left(\frac{R_{A}}{r}\right)^{4}\left(\frac{1+\lambda
r}{1+\lambda R_{A}}\right)^{2}e^{-2\lambda (r-R_{A})},\quad r>R_A
\label{eq:VLi}
\end{equation}
where
\begin{equation}
V_0=\frac{g^{2}}{3m_{n}} \frac{\vec{\alpha}^{2}}{3}\eta^{2}\simeq
61.3 \,\mathrm{MeV} \label{Vo}
\end{equation}
with
\begin{equation}
\eta=8\pi\sqrt{\frac{\rho}{2\omega_{0}}}j_{1}(qR_{A}) \label{an}
\end{equation}
Writing for $l=0$
\begin{equation}
\psi(r)=\frac{1}{r}\chi(r) \label{reduwave}
\end{equation}
and assuming for the interaction potential the following
approximate expression
\begin{equation}
V_{ap}(r)=-V_{0}e^{-\mu (r-R_A)},\quad r>R_A \label{appot}
\end{equation}
with $\mu=1.8$ fm${}^{-1}$, calculated by interpolation of the
potential (\ref{appot}) with the exact potential (\ref{eq:VLi}),
we can introduce the new variable
\begin{equation}
y=e^{-\frac{\mu}{2} (r-R_A)} \label{newvar}
\end{equation}
and rewrite the radial part of the Schr\"{o}dinger equation as
\begin{equation}
\frac{d^{2}\chi}{dy^{2}}+\frac{1}{y}\frac{d\chi}{dy}+\left(c^{2}
-\frac{q^{2}}{y^{2}}\right)\chi=0 \label{reduschroedap}
\end{equation}
with the abbreviations
\begin{equation}
c^{2}=16\frac{m_n V_{0}}{\mu^{2}}, \quad q^{2}=-16\frac{m_n
E_{2n}}{\mu^{2}} \label{abb2}
\end{equation}
The equation (\ref{reduschroedap}) is a differential Bessel
equation, whose general solution is given by
\begin{equation}
\chi (y)=C_{1}J_{q}(cy)+C_{2}J_{-q}(cy).
 \label{solap}
\end{equation}
By Eq. (\ref{newvar}) we have that for $y=0$, $r\rightarrow
\infty$ and $\chi$ must vanish. Therefore $C_{2}=0$ and the
reduced wave function becomes
\begin{equation}
\chi(r)=C_{1} J_{q}\left[ce^{-\frac{\mu}{2} (r-R_A)}\right]
\label{new sol}
\end{equation}
Due to the Pauli principle between the extra and the core
nucleons, we require $\chi (R_A)=0$ i.e.
\begin{equation}
J_{q}\left(c\right)=0\label{ipj}
\end{equation}
stipulating that the surface of the nucleus acts as a infinite
potential barrier.

Using Eq. (\ref{abb2}) and (\ref{ipj}) and the above numerical
values for $\mu$ and $V_{0}$, we can calculate the energy
eigenvalue of the bound state and we obtain
\begin{equation}
E_{2n}\simeq\, 300 \,\mathrm{keV}. \label{anlyt-ene}
\end{equation}
to be compared with the experimental value $E=294\pm 30$ keV
\cite{Tani}. By means of the derived wave function the root mean
square radius of the halo is given by
\begin{equation}
r_{h}=\sqrt{\int^{+\infty}_{R_{A}}r^{2}\chi^2(r)dr}\simeq
7.0\,\mathrm{fm} \label{root2}
\end{equation}
so that the root mean square radius of the total system is
\begin{equation}
r_{RMS}=\sqrt{\frac{A_{c}}{A}r_{c}^2+\frac{2}{A}r_{h}^2}\simeq 3.6
\,\mathrm{fm} \label{rms2}
\end{equation}
where $r_{c}$ is $rms$ radius of core nucleus. The experimental
value of the above quantity is $3.55 \pm 0.10$ fm \cite{kaili}.

Let us now compare the potential obtained for one neutron and the
above one just calculated for the system of the two neutrons. We
have
\begin{equation}
    \frac{(V_0)_{n}}{(V_0)_{2n}}\propto \frac{(\tau \cdot
    \sigma)^2}{(\tau_{1} \cdot \sigma_{1}+\tau_{2} \cdot
    \sigma_{2})^{2}}=\frac{3} {4}
    \label{confrpot}
    \end{equation}
The limiting value of $V_0$ is obtained from (\ref{ipj}) in the
limit of vanishing energy $q=0$ and is given by
\begin{equation}
    (V_0)_{min}=\frac{(2.4 \mu)^2} {16 m_n}\simeq 48.5\, \mathrm{MeV}
    \label{limval}
    \end{equation}
From (\ref{confrpot}) we obtain a smaller value
\begin{equation}
    (V_0)_{n}\simeq 46.5 \, \mathrm{MeV}
    \label{confrpot2}
    \end{equation}
This fact points out that the ${}^{10} Li$ is unbound as expected.

\section{Conclusions}

The qualitative difference between the standard approaches and our
calculation is clear and can be easily understood. In ordinary
potential models the neutrons are loosely bound to an inert core
and occupy the (possible) vacant state of the nuclear potential
and the pairing between the neutrons in two-neutron halo plays a
crucial role in their stability. In our approach the extra
neutrons are localized outside the core through the virtual
interaction with the evanescent $\pi$-wave. On the other hand, if
$\pi$-condensation does occur, like it happens in our model, an
interaction between two neutrons and the evanescent $\pi$-wave
become possible, leading to a halo nucleus. In this way, we are
able to reproduce the experimental results for the two-neutron
separation energy and root mean square radius of ${}^{11} Li$ and
the fact that the ${}^{10} Li$ is unbound.

With this work we have taken only the first step in a research
program aimed at analyzing the consequences of the Coherent
Nucleus Theory on the structure of halo nuclei and leave for a
future publication a more detailed investigation of the
correlations in two neutron halos.


\end{document}